
\font\bigbf=cmbx10 scaled 1400
\hfuzz 25pt

\hsize=6.1truein
\hoffset=0.2truein

\def\go{\mathrel{\raise.3ex\hbox{$>$}\mkern-14mu
             \lower0.6ex\hbox{$\sim$}}}
\def\lo{\mathrel{\raise.3ex\hbox{$<$}\mkern-14mu
             \lower0.6ex\hbox{$\sim$}}}
\def\cC {{\cal C}}
\def\Ib5{{\hbox{\rlap{\hbox{\raise.27ex\hbox{-}}}I}}^{(5)}}

\centerline{\bigbf GRAVITATIONAL RADIATION FROM RAPIDLY}
\medskip
\centerline{\bigbf ROTATING NASCENT NEUTRON STARS}
\bigskip

\centerline{
DONG LAI$\,$\footnote{$^1$}
{Address after September 1, 1994:
Theoretical Astrophysics, California Institute of Technology,
Pasadena, CA 91125}
and $\,$STUART L. SHAPIRO$\,$\footnote{$^2$}
{Department of Astronomy and Physics, Cornell University.}}

\medskip
\centerline{\it Center for Radiophysics and Space Research,
Space Sciences Building}
\centerline{\it Cornell University, Ithaca, NY 14853}

\medskip
\centerline{E-mail: dong@astrosun.tn.cornell.edu ~~~
                shapiro@astrosun.tn.cornell.edu}

\bigskip
\bigskip
\centerline{\bf ABSTRACT}
\medskip

We study the secular evolution and gravitational wave signature of
a newly-formed, rapidly rotating neutron star.  The neutron star may
arise from core collapse in a massive star or from the accretion-induced
collapse of a white dwarf. After a brief dynamical phase, the nascent
neutron star settles into an axisymmetric, secularly unstable equilibrium
state. Gravitational radiation drives the star to a nonaxisymmetric,
stationary equilibrium configuration via the bar-mode instability.
The emitted quasi-periodic gravitational waves have a unique signature:
the wave frequency sweeps downward from a few hundred Hertz to zero, while
the wave amplitude increases from zero to a maximum and then decays back to
zero. Such a wave signal could be detected by broad-band gravitational
wave interferometers currently being constructed.

We also characterize two other types of gravitational wave signals
that could arise in principle from a rapidly rotating, secularly unstable
neutron star: a high-frequency ($f\go 1000$ Hz) wave which increases
the pattern-speed of the star, and a wave that actually increases the
angular momentum of the star.


\bigskip
\bigskip
{\it Subject headings:} hydrodynamics --- instabilities --- stars:
core collapse, supernova --- stars: neutron ---
radiation mechanisms: gravitational

\bigskip
\bigskip
\centerline{\it Submitted to ApJ, May 1994}
\medskip

\vfil\eject
\bigskip
\centerline{\bf 1. INTRODUCTION}
\medskip

The advent of the new generation of gravitational wave detectors,
such as the Laser Interferometer Gravitational Wave Observatory (LIGO;
see Abramovici et al.~1992) and its French-Italian counterpart VIRGO
(Bradaschia et al.~1990), stimulates a renewed interest in astrophysical
sources of gravitational waves (Thorne 1987, 1992).
Of special importance are those sources that radiate in LIGO's
sensitivity band, $10-1000$ Hz.
In this paper we study possible unique gravitational wave signals
with frequency in this range from newly-formed neutron stars.

Gravitational collapse leading to the formation of a neutron star
has long been considered an observable source of gravitational radiation.
The event rate of type II supernovae is estimated to be
several per year out to a distance of $10$ Mpc, the distance to
the Virgo cluster of galaxies (van den Bergh \& Tammann 1991). The rate of
electromagnetically ``quiet'' gravitational
collapses (such as accretion-induced collapse of white dwarfs)
may be as high as one per galaxy every one to five years
(Bahcall \& Piran 1983, Blair 1989).
However, the strengths of gravitational wave signals
from core collapse are still highly uncertain because
they depend crucially on the unknown asymmetry
of the collapse. If the collapse is axisymmetric, the energy radiated during
the implosion phase of the collapse is at most $\sim 5\times 10^{-8}Mc^2$
(Finn 1991, M\"onchmeyer et al.~1991).
Such a supernova would not be detectable, even by the advanced LIGO detectors,
for sources beyond our galaxy and the Magellanic clouds.
However, it is widely expected that if the collapsing and bouncing core
rotates so rapidly that it becomes nonaxisymmetric, the efficiency
of producing gravitational wave can be much higher.
So far, nonaxisymmetric core collapse has not been studied
except for the calculations based on classical homogeneous ellipsoid models
(see Saenz \& Shapiro 1981 and references therein).

It is well known that nonaxisymmetric instabilities can develop
in rapidly rotating fluid bodies when the ratio $\beta\equiv T/|W|$
of the rotational energy $T$ to the gravitational potential energy $W$
is sufficiently large (Chandrasekhar 1969; hereafter referred as Ch69).
A star becomes dynamically unstable when $\beta >\beta_{dyn}\simeq 0.27$,
while its $l=m=2$ ``bar'' mode oscillations become secularly unstable
due to viscosity (Roberts \& Stewartson 1963)
or gravitational radiation reaction (Chandrasekhar 1970a) when
$\beta >\beta_{sec}\simeq 0.14$.
The gravitational radiation (GR) instability for higher-order modes ($m>2$)
can set in at even smaller value of $\beta$
(Friedman \& Schutz 1978, Comins 1979a,b).
When the degenerate core of a massive star or when a white dwarf with mass
exceeding the Chandrasekhar limit
collapses to form a neutron star, the value of $\beta$ can increases
significantly. Assuming angular momentum conservation,
the collapsed star would have
$${T\over |W|}={R_i\over R}\left({T\over |W|}\right)_i,$$
where $R_i$ and $R$ are the radius of the core before and after
collapse, and $\beta_i$ refers to the precollapse initial value. Typically,
$R_i\sim 1500$ km for a white dwarf near the Chandrasekhar mass,
and $R\sim 10$ km for a neutron star; hence $R_i/R\sim 10^2$.
Thus the nascent neutron star can be rapidly rotating even if the
initial precollapse core is not.

While there are no direct observations of the rotation of precollapse
cores --- they are hidden inside the envelopes of giant stars ---
the situation for the accretion-induced collapse of white dwarf
and for the collapse of merging white dwarf binaries may be clearer.
In both cases, the precollapse
white dwarfs can attain sufficient angular momentum from the accreting
gas or from the initial binary orbits.
Although most millisecond pulsars are thought to have been spun up
by accretion, it has been suggested that some millisecond pulsars
originate from the collapse of white dwarfs in binary systems
(Michel 1987, van den Heuvel 1987, Bailyn \& Grindlay 1990).
The initial spins of ordinary radio pulsars are not well constrained
by the pulsar statistics; but it is possible that some radio pulsars
are created with high initial spin.
Although the physical mechanism that gives rise to
the high velocity ($\sim 1000$ km/s) of some radio pulsars such as
PSR 2224+65 (Cordes, Romani \& Lundgren 1993) is not completely clear
(see Radhakrishnan 1992 for a review),
it is conceivable that rapid rotation and asymmetry of the collapsed core
play an important role.
Finally, there may already be some indirect observational
evidence for asymmetric core collapse:
the spectropolarimetry of SN1987a (Cropper et al.~1988)
and SN1993J (Trammell et al.~1993), as well as emission line
profiles in the spectra of SN1993J (Spyromilio 1994), indicate an asymmetric
atmosphere in both supernovae.

A new-born neutron star can be secularly unstable but
dynamically stable ($\beta_{sec}<\beta<\beta_{dyn}$) only
if the rotation rate of the precollapse core lies in a narrow range.
Thus it is more likely that the core becomes dynamically unstable
($\beta>\beta_{dyn}$) following collapse,
provided that the initial $\beta_i$ is not too small.
The fate of a dynamically unstable rotating star has been studied numerically
by Durisen et al (1986), Williams \& Tohline (1988), and more
recently by Houser, Centrella \& Smith (1994), who have also calculated the
resulting gravitational waves due to the dynamical instability. According to
Houser et al (1994), the instability develops as follows:
after shedding mass and angular momentum in the form
of two tailing spiral arms, which then expand and merge, causing shock
dissipation, the system evolves on a dynamical timescale
toward a nearly axisymmetric equilibrium state.
Its $\beta$ falls below the dynamical stability limit $\beta_{dyn}$,
but remains much larger than
the secular stability limit $\beta_{sec}$. Such a post-collapse,
axisymmetric core will evolve into a nonaxisymmetric configuration
on a secular dissipation timescale. The main purpose of our
paper is to study the growth of this secular
instability (\S 2) and to characterize the
resulting gravitational wave form (\S 3). We follow the secular evolution
using a compressible ellipsoid model, which we have recently developed
to study the equilibrium and dynamics of rotating stars and binary
systems (Lai, Rasio \& Shapiro 1993,~1994, hereafter referred as Paper I and
Paper II). We shall see that for the typical neutron
star with mass $M\sim 1.4 M_\odot$ and radius $R\sim 10$ km,
the gravitational wave amplitude increases from zero to a maximum
and then decreases to zero again, while
the wave frequency sweeps from a few hundred Hertz down to zero (Fig.~4).
Such a gravitational wave signal can in principle be detected by LIGO
(\S 3.3, especially Fig.~6).

Another purpose of our paper is to characterize {\it all\/} types of
gravitational waveforms one can expect from a rotating star
modeled as a compressible fluid ellipsoid.
The secular evolution of homogeneous, incompressible ellipsoidal figures
has been studied before (Miller 1974, Detweiler \& Lindblom 1977), but the
waveforms have not been considered or characterized.
In addition to the low-frequency
wave considered in \S 3, we find two other types of
wave forms: a high-frequency wave with decreasing amplitude and increasing
frequency, and a peculiar wave which increases the angular momentum
of the star. These are discussed in \S 4.

The fluid viscosity can have some effects on the secular instability
and the evolution. These effects are discussed in \S 5.
A summary including a discussion of the main uncertainties is given in
in \S 6.

We use units such that $G=c=1$ throughout the paper.

\bigskip
\centerline{\bf 2. GRAVITATIONAL RADIATION INSTABILITY IN}
\smallskip
\centerline{\bf ROTATING STARS}
\medskip

Chandrasekhar (1970a) first discovered that
GR reaction can induce a nonaxisymmetric secular instability in a
uniformly rotating, incompressible Maclaurin spheroid.
He found that the $l=m=2$ ``bar'' mode
(with angular behavior of the form $e^{\pm im\phi}$)
becomes unstable beyond the point at which the nonaxisymmetric Jacobi
sequence or Dedekind sequence bifurcates from the axisymmetric
Maclaurin sequence (Ch69). At the bifurcation point,
the ratio of the kinetic energy and the
absolute value of the potential energy $\beta\equiv T/|W|$ equals
$\beta_{sec}=0.1375$.

We have re-analyzed this bar-mode instability
induced by GR as well as by viscosity in {\it compressible\/} rotating stars
(Appendix A). Our analysis is based on the equilibrium
and dynamical ellipsoid model developed in Papers I-II,
in which a rotating star is treated as a self-similar ellipsoid
with polytropic density and linear velocity profiles.
We have extended many known results concerning
incompressible ellipsoidal figures (Ch69) to their compressible analogs.
In particular, we find that, just as
in the incompressible case, the secular instability sets in at
the bifurcation point, where $\beta=\beta_{sec}=0.1375$ independent of the
polytropic index $n$. Similarly, a dynamical bar-mode instability sets
in at $\beta=\beta_{dyn}=0.2738$. While for strictly uniformly
rotating stars, mass-shedding can occur before the point of
bifurcation for $n>0.808$ (James 1964; see Tassoul 1978 for review),
a slight amount of differential rotation can
in principle inhibit mass-shedding without
changing the global structure of the star.
Indeed, axisymmetric models of differentially rotating polytropes which
retain the global properties of Maclaurin spheroids have been constructed
by Bodenheimer \& Ostriker (1973). Our finding that the secular
and dynamical stability limits do not
depend sensitively on the polytropic index $n$ as well as differential
rotation law is also corroborated by earlier numerical studies
(Ostriker \& Bodenheimer 1973; see also Managan 1985;
Imamura, Friedman, \& Durisen 1985; Ipser \& Lindblom 1990).

Along an equilibrium Maclaurin sequence parametrized by the eccentricity $e$,
the ratio $\beta$ and the dimensionless angular velocity $\hat\Omega_M$ are
given by
$$\beta={3\over 2e^2}\left[1-{e(1-e^2)^{1/2}\over\sin^{-1}e}\right]-1,
\eqno(2.1)$$
and
$$\hat\Omega_M^2\equiv q_n{\Omega_M^2\over\pi\bar\rho}
=2\left[{(1-e^2)^{1/2}\over e^3}(3-2e^2)\sin^{-1}e
-{3(1-e^2)\over e^2}\right],\eqno(2.2)$$
where $q_n=(1-n/5)\kappa_n$, $\kappa_n\le 1$ is a constant of order unity
which depends only on $n$ (see Table I in Paper I), and
$\bar\rho=3M/(4\pi R^3)$ is the mean density.
Equations (2.1)-(2.2) are the same as the expressions for
incompressible Maclaurin spheroids (Ch69),
but they are also valid for compressible configurations within our
approximation.
Because of the centrifugal support, the size of the star increases as the
rotation rate increases.
The mean radius of the rotating star $R\equiv (a_1a_2a_3)^{1/3}$ is given by
$$R=R_o\left[{\sin^{-1}e\over e}(1-e^2)^{1/6}(1-\beta)\right]^{-n/(3-n)},
\eqno(2.3)$$
where $R_o$ is the radius of the nonrotating star with the same mass.
For a given equation of state and mass, $R_o$ is uniquely determined.
In a rotating frame with angular frequency $\Omega$, our analysis
(Appendix A) yields the bar-mode frequency
$$\sigma_\pm(\Omega)=(2\Omega-\Omega_M)\pm\sigma_o,\eqno(2.4)$$
where
$$\sigma_o=\left({4\over q_n}\pi \bar\rho B_{11}-\Omega_M^2\right)^{1/2},
\eqno(2.5)$$
and $B_{11}$ is the index symbol as defined in Ch69
(Chap.~3), and depends only on the axis ratios.
Gravitational radiation de-stablizes the Dedekind-mode (the mode with plus
sign in eq.~[2.4]) when $\hat\Omega_M^2>2B_{11}$,
or, equivalently, when $\beta>\beta_{sec}$. The growth time $\tau_{GW}$ for
the secular instability is given by
$$\eqalign{
\tau_{GW}=&{25\over 2\kappa_nMa_1^2}
{\sigma_o\over (\Omega_M-\sigma_o)^5}\cr
=&0.084\,M_{1.4}^{-3}R_{10}^{4}\kappa_n\left(1-{n\over 5}\right)^2
\left[\left({R\over R_o}\right)^4{(1-e^2)^{1/3}
\hat\sigma_o\over (\hat\Omega_M-\hat\sigma_o)^5}\right]~({\rm s}),\cr
}\eqno(2.6)$$
where $M_{1.4}\equiv M/(1.4M_\odot)$, $R_{10}\equiv R_o/(10~{\rm km})$,
and $\hat\sigma_o\equiv q_n^{1/2}\sigma_o/(\pi G\bar\rho)^{1/2}
=(4B_{11}-\hat\Omega_M^2)^{1/2}$ (eq.~[2.5]).
For $0<\beta-\beta_{sec}<<1$, equation (2.6) can be approximately written as
$$\tau_{GW}\simeq 2\times 10^{-5}\,M_{1.4}^{-3}R_{10}^{4}
(\beta-\beta_{sec})^{-5}~({\rm s}),\eqno(2.7)$$
Figure~1 shows the $\tau_{GW}$ as a function of $\beta=T/|W|$ for
$n=0,~0.5,~1$.
We see that for large $\beta$, the growth time $\tau_{GW}$ is larger for
more compressible stars because of the rotation-induced
expansion of the star (eq.~[2.3]).
However, for $n\sim 0.5-1$, typical of neutron stars, $\tau_{GW}$
does not depend sensitively on the polytropic index $n$.
Thus for a neutron star with $M=1.4M_\odot$, $R_o=10$ km, we have
$\tau_{GW}\simeq 7\times 10^4$ s for $\beta=0.15$,
$\tau_{GW}\simeq 20$ s for $\beta=0.20$, and
$\tau_{GW}\simeq 1$ s for $\beta=0.24$.

Higher-order ($m>2$) modes can become unstable at a smaller $\beta$.
In fact, gravitational radiation tends to make all rotating star
unstable (Friedman \& Schutz 1978). However, the growth time
of this gravitational radiation instability increases
rapidly as $m$ increases (Comins 1979a,b).
Moreover, viscosity tends to counteract this
instability (Lindblom \& Detweiler 1977), and the
stablizing effect becomes more efficient for higher order modes.
Thus we will not discuss these high-order instabilities
for $\beta<\beta_{sec}$; the consequences of these instabilities
are not as clear as that of the bar-mode instability.
Note that for $\beta$ just slightly larger than $\beta_{sec}$,
the growth time for the $m=2$ mode instability is longer
than that of a $m>2$ mode (as seen in Fig.~1, where the result for
the $m=3$ mode is from Comins 1979b, for $n=0$).
Thus our discussion below about the evolution will not be applicable to
those very small values of $\beta-\beta_{sec}$. However, for $\beta>0.17$,
the growth time for the bar-mode is much shorter
(by more than a factor of $10$) than that of the $m=3$ mode
Therefore, for these values of $\beta$,
the bar-mode instability is much stronger
than the higher-order modes, and our discussion of the secular evolution is
valid.

Note that we have normalized the mass and radius to those of a cold,
nonrotating neutron star. Following its implosion, which takes $\sim 1$ s,
the $1.4M_\odot$ collapsing core settles
into hydrostatic equilibrium within a few
milliseconds. This hot protoneutron star does not resemble
a usual cold neutron star, as it consists of an unshocked cold inner core
with radius $\sim 10$ km surrounded by a shocked hot outer core (mantle)
which extends to $\sim 100$ km.
The protoneutron star evolves quasi-statically via neutrino cooling.
However, within one second,
the outer mantle cools and contracts significantly,
and the protoneutron star (with mass $\sim 1.4M_\odot$)
approaches a canonical radius of $10$ km. By this time, the hydrostatic
structure of the neutron star more or less resembles that of a cold
neutron star, although the star will remain
very hot ($T\go 10^{10}$ K) for more than a few tens of seconds
(Burrows \& Lattimer 1986, Burrows 1990).
Since the nonaxisymmetric instability growth time
is typically larger than a second, it is appropriate to use
the canonical radius of a cold neutron star (but see \S 6).

\bigskip
\centerline{\bf 3. EVOLUTION AND GRAVITATIONAL WAVEFORM}
\medskip

Here we discuss the evolution of a rotating star modeled
as a non-axisymmetric ellipsoid (i.e., a Riemann-S ellipsoid
in the language of Ch69 and Paper I) under the influence of
gravitational radiation reaction and zero viscosity.

\bigskip
\centerline{\bf 3.1 Secular Evolution}
\medskip

A secularly unstable Maclaurin spheroid will evolve away
from the axisymmetric configuration due to gravitational radiation.
The evolution can in principle be studied using the full
dynamical equations of ellipsoidal figures (Ch69, Miller 1974, Paper II)
including gravitational radiation reaction.
However, since the growth time of the instability $\tau_{GW}$ is generally
much longer than the dynamical time of the star, the evolution is
quasi-static, i.e., the star will evolve along an equilibrium
sequence of Riemann-S ellipsoids (Ch69, Paper I). A general
Riemann-S ellipsoid is characterized by the angular velocity
${\bf\Omega}=\Omega{\bf e}_z$ {\it of the ellipsoidal figure} (the pattern
speed) about a principal axis
and the internal motion of the fluid with uniform vorticity
${\bf\zeta}=\zeta{\bf e}_z$ along the same axis
(in the frame corotating with the figure).
The vorticity in the inertial frame is then $\zeta^{(0)}=2\Omega+\zeta$.
Since the gravitational radiation reaction acts like a
potential force, the fluid circulation $C$ along the
equator of the star is conserved in the absence of
viscosity (Miller 1974; Paper II). Following Paper I (\S 5.1) we write
$$\cC\equiv \biggl (-{1\over5\pi}\kappa_nM\biggr)\,C
=\biggl (-{1\over5\pi}\kappa_nM\biggr)\,\pi a_1a_2\, {\zeta}^{(0)}
= I\Lambda-{2\over5}\kappa_nMa_1a_2\Omega,\eqno(3.1)$$
where $I=\kappa_nM(a_1^2+a_2^2)/5$ is the moment of inertia, and
$\Lambda=-a_1a_2\zeta/(a_1^2+a_2^2)$.
The quantity $\cC$ has the dimension of angular momentum but is proportional
to the conserved circulation $C\equiv \pi a_1a_2 {\zeta}^{(0)}$.
We refer to $\cC$ itself as the circulation for convenience.
Note that in the same notation, the angular momentum of the
ellipsoid is
$$J=I\Omega-{2\over5}\kappa_nMa_1a_2\Lambda.\eqno(3.2)$$

Thus, under the influence of gravitational radiation
reaction, and when the viscous dissipation is negligible,
a secularly unstable Maclaurin spheroid will evolve along a
constant-$\cC$ equilibrium sequence, and proceed ultimately toward
a Dedekind ellipsoid, for which $\Omega=0$ (a ``stationary football'').
The evolution track can be best illustrated by the diagram
shown in Figure 2, which depicts the equilibrium energies for
various constant$-\cC$ sequences as functions of the axis ratio $a_2/a_1$,
all for $n=1$. For comparison, the Jacobi and the Dedekind sequences are
also shown, while the Maclaurin sequence corresponds to the vertical line
with $a_2/a_1=1$.
A given Maclaurin spheroid has a unique value of
$\cC$, but two different constant$-\cC$ sequences branch off:
one sequence is {\it Jacobi-like}, with $|\zeta|<2|\Omega|$,
the other is {\it Dedekind-like}, with $|\zeta|>2|\Omega|$.
Clearly, the endpoint of the evolution depends on the value $|\cC|$
of the initial configuration.
For $|\cC|<|\cC_{sec}|$, where $\cC_{sec}$ is the circulation
of a Maclaurin spheroid
at the bifurcation point, the Maclaurin spheroid is the lowest energy
terminal state toward which both Jacobi-like and Dedekind like configurations
evolve; for $|\cC|>|\cC_{sec}|$, on the other hand,
the terminal state is a Dedekind ellipsoid. A Jacobi-like
configuration with $|\cC|>|\cC_{sec}|$ first evolves
toward an axisymmetric, Maclaurin state, and then continues through a
Dedekind-like phase toward a Dedekind ellipsoid.
The value of $|\cC_{sec}|=J_{sec}$ is given by
$$J_{sec}=0.304\,\left({\kappa_n\over 1-n/5}\right)^{1/2}(M^3R)^{1/2},
\eqno(3.3)$$
(see eq.~[3.27] and Table II in Paper I), where $R$ is
given by equation (2.3).

Now let us focus on the evolution of a secularly unstable Maclaurin spheroid
toward a Dedekind ellipsoid, corresponding to region I in Figure~2.
The initial Maclaurin spheroid has angular momentum $J_i=|\cC_i|$. The final
Dedekind state has $J_f=|\cC_i|[2a_1a_2/(a_1^2+a_2^2)]_f$ (cf.
eqs.~[3.1]-[3.2] with $\Omega=0$), the total angular
momentum radiated in gravitational wave is given by
$$\Delta J=J_i\left[{(a_1-a_2)^2\over a_1^2+a_2^2}\right]_f,\eqno(3.4)$$
where the subscript ``f'' stands for the final Dedekind configuration.
The total energy $\Delta E$
emitted as gravitational radiational wave for a secularly
unstable Maclaurin spheroid can be calculated by comparing the
difference in equilibrium energy between the Maclaurin spheroid and the
Dedekind ellipsoid with the same $\cC$. Figure 3 shows the dependence of
$\Delta E$ as a function of $\beta=T/|W|$ of the initial Maclaurin spheroid.
We see that $\Delta E$ does not depend sensitively the
compressibility for $n\lo 1$. Thus the energy radiated in gravitational
wave during the Maclaurin-Dedekind transition can be as large as
$0.02M^2/R_o$, or $\Delta E\lo 4\times 10^{-3}M$ for a typical
neutron star radius ($R_o/M\simeq 5$). This is much larger than
the energy radiated ($\lo 5\times 10^{-8}M$) in a axisymmetric
collapse preceding the neutron star formation
(Finn 1991, M\"onchmeyer et al.~1991).

\bigskip
\centerline{\bf 3.2 Gravitational Wave Amplitude}
\medskip

The gravitational wave emitted during the Maclaurin-Dedekind evolution
has certain unique characteristics. At the beginning of the
evolution, the wave amplitude is small since
the star is still nearly axisymmetric. Near the end of the evolution,
as the star approaches a Dedekind configuration asymptotically,
the wave amplitude is small again because the angular velocity $\Omega$ of
the figure becomes decreasingly small. Thus we expect
{\it a non-monotonic
behavior of the gravitational wave amplitude during such a evolution,
while the wave frequency decreases monotonically.\/}

In the weak-field, slow-motion limit, the gravitational radiation
can be calculated by the standard quadrupole formula (see, e.g.,
Misner, Thorne \& Wheeler 1970).
The rates of energy loss and angular momentum loss are given by
$$\left({dE\over dt}\right)_{GW}=\Omega\left({dJ\over dt}\right)_{GW}
=-{32\over 5}\Omega^6(I_{11}-I_{22})^2,
\eqno(3.5)$$
where $I_{ii}=\kappa_nMa_i^2/5$ are the components of the star's
quadrupole moment along the principal axes in its equatorial plane.
At a distance $D$ from the source,
the two polarization components of the waveforms are
$$\eqalign{
h_{+} &= -{2\over D}\Omega^2 (I_{11}-I_{22})
\cos\Phi (1+\cos^2\theta),\cr
h_{\times} &=-{4\over D}\Omega^2 (I_{11}-I_{22})
\sin\Phi\cos\theta,\cr
}\eqno(3.6)$$
where $\theta$ is the angle between the rotation axis of the star and
line of sight from the earth, and
$\Phi\equiv 2\int^t\Omega dt$ is twice the orbital phase.
The wave is quasi-periodic,
with frequency $f=\Omega/\pi$ and amplitude (for $\theta=0$)
$$h={4\over D}\Omega^2 (I_{11}-I_{22}).\eqno(3.7)$$

Typical wave amplitudes as a function of frequency
are shown in Figure 4, for polytropic index
$n=0.5$ and $1$. The initial Maclaurin spheroids
have $\beta=0.2$ and $\beta=0.24$, respectively.
The wave sweeps with time
from high frequency to low, that is, from right to left in Figure 4.
Also shown in the figure are the
timescale of the evolution as measured by $|dt/d\ln f|=|f/\dot f|$,
and the number of cycles gravitational wave spent near frequency $f$
$$\left|{dN\over d\ln f}\right|=\left|{f^2\over\dot f}\right|
={\Omega^2\over\pi}\left|{dE\over d\Omega}\right|_{eq}
\left|\left({dE\over dt}\right)_{GW}^{-1}\right|,
\eqno(3.8)$$
where $dE/d\Omega$ is evaluated along an equilibrium sequence with
constant $\cC$.
The number of cycles near the frequency at which the wave amplitude is
maximum depends sensitively on the initial value of $\beta$,
and to a less extent, on the polytropic index. But even
for the initial $\beta$ close to the dynamical limit $\beta_{dyn}$,
there are still $\go 10^3$ cycles when the
wave amplitude is close to the maximum.
Note that at the beginning of the evolution ($f\rightarrow f_{max}$),
the timescale approaches infinity. This is because we have assumed exact
equilibrium models, and a {\it perfect\/} Maclaurin spheroid does not
evolve at all. In reality, a small perturbation will make the Maclaurin
spheroid evolve on the order of a few growth time $\tau_{GW}$,
as determined in \S 2. This time is of the same order
of magnitude as the time the system spends near the maximum amplitude phase.

It is important to note that {\it the wave frequency is determined by the
pattern speed $\Omega$ of the ellipsoidal figure, and this frequency
can be much lower than the rotational frequency of the
secularly unstable star\/}. During the evolution, the
wave frequency decreases monotonically from an initial value to zero.
Consider the maximum frequency of the wave, corresponding to
the frequency near the beginning of the evolution,
when the star is still close
to a Maclaurin spheroid. The bar-mode oscillation frequency as viewed in a
general rotating frame is given by equation (2.4),
and gravitational radiation destablizes the Dedekind-mode.
In a particular reference frame, with angular velocity $\Omega_{max}$,
the Dedekind-mode can be neutralized, i.e., $\sigma_{+}(\Omega_{max})=0$.
It is along the axes of this rotating frame that the Maclaurin
spheroid starts deforming into a nonaxisymmetric shape.
Therefore, as the Maclaurin spheroid begins to evolve
due to gravitational radiation, the pattern speed $\Omega_{max}$ is given by
$$\Omega_{max}={1\over 2}(\Omega_M-\sigma_o),\eqno(3.9)$$
which is always smaller than the initial rotation rate
$\Omega_M$ of the star.
Clearly, at the bifurcation point, $\Omega_{max}=0$
(i.e., at the bifurcation point, the oscillation mode
is neutralized in the inertial frame).
The angular frequency $\Omega_{max}$ is also one of the two values
assigned to a Maclaurin spheroid if it is to be considered as
a limiting member of Riemann ellipsoids (Ch69, \S 48).
Equation (3.9) is simply half of the absolute value
of the mode frequency $\sigma_{+}(0)$ in the inertial frame.
The maximum frequency of the gravitational wave during the Maclaurin-Dedekind
evolution is then
$$\eqalign{
f_{max} &={\Omega_{max}\over\pi}
={\sqrt{3}\over 4\pi}\left({GM\over R_o^3}\right)^{1/2}
q_n^{-1/2}\left({R_o\over R}\right)^{3/2}
(\hat\Omega_M-\hat\sigma_o)\cr
&=1.88\times 10^3M_{1.4}^{1/2}R_{10}^{-3/2}q_n^{-1/2}
\left({R_o\over R}\right)^{3/2}
\left[\hat\Omega_M-(4B_{11}-\hat\Omega_M^2)^{1/2}\right]~({\rm Hz}).\cr
}\eqno(3.10)$$
Figure 5 shows $f_{max}$ as a function of initial $\beta$ for $n=0,~0.5$ and
$1$. For a given $\beta$, the maximum frequency is smaller for larger $n$,
as a result of the rotational expansion factor $R/R_o$.
This dependence on $n$ is
significant only for larger $\beta$ and $n\go 1$.
For comparison, we also show in Figure 5 twice of the rotational frequency of
the Maclaurin spheroid $2f_{M}=\Omega_M/\pi$. This would be the
frequency of the gravitational wave emitted if the rapidly rotating star
were deformed slightly away from axisymmetry by external stress
(e.g., a mountain on the neutron star). This frequency is larger than
$1000$ Hz if star rotates sufficiently rapidly to be secularly unstable.

\bigskip
\centerline{\bf 3.3 Characteristic Wave Amplitude}
\medskip

Thorne (1987) has emphasized the importance of the signal-to-noise ratio of
a given source when considering its detectability.
For a broad-band detector such as LIGO, the best signal-to-noise
ratio will be obtained by matched filtering of the data,
$S/N\simeq h_c/h_{rms}$, where $h_{rms}$ is the interferometer's root
mean square noise (Abramovici et al 1992),
and $h_c$ is the characteristic amplitude of the source.
For the quasi-static evolution discussed in \S 3.2, $h_c$ is given by
\footnote{$^3$}{More precisely, one needs to average over the source
direction to obtain $h_c$ (Thorne 1987), but this only introduces
a factor $\sqrt{4/5}$ in eq.~(3.11).}
$$h_c\equiv h\left|{dN\over d\ln f}\right|^{1/2}
={1\over D}\left({5\over 2\pi}\left|{dE\over d\Omega}\right|_{eq}\right)^{1/2}
={M\over D}\left({R_o\over M}\right)^{1/4}
\left({5\over 2\pi}\left|{d\bar E\over d\bar\Omega}\right|_{eq}\right)^{1/2}
\eqno(3.11)$$
where $\bar E=E/(M^2/R_o)$, $\bar\Omega=\Omega/(M/R_o^3)^{1/2}$, and
we have used equations (3.7)-(3.8).

In Figure 6, we show $h_c$ for the four wave amplitudes depicted in Figure 4,
and compare them with the rms noise $h_{rms}$
and the sensitivity to bursts
$h_{sb}=11 h_{rms}$ of the advanced LIGO detector (Abramovici et al.~1992).
 From our numerical calculations, we find $(d\bar E/d\bar\Omega)_{eq}\propto
\bar\Omega$ to a good approximation. Thus the characteristic
amplitude of gravitational wave during the evolution from a
Maclaurin spheroid to a Dedekind Ellipsoid is given by
$$h_c\simeq 6.0\times 10^{-23}\left({30\,{\rm Mpc}\over D}\right)
M_{1.4}^{3/4}R_{10}^{1/4}f^{1/2},\eqno(3.12)$$
where $f$ is the wave frequency in Hertz.
This expression is accurate to within $\sim 20\%$ to all relevant
values of $\beta$ and $n$ ($0.16<\beta<0.27$, $0\le n<1.5$).
We see clearly from Figure 6 that {\it the gravitational waves from
the non-axisymmetric evolution of rapidly rotating neutron stars
should be detectable with high confidence ($S/N>11$)
out to the distance of $\sim 140$ Mpc
by the advanced LIGO detector at $\sim 100$ Hz.\/}
Valuable information about the masses and radii, as well as
the rotation rates of newly-formed neutron stars may be
gained from the unique wave signatures (Figs.~4-5).

\bigskip
\centerline{\bf 4. OTHER TYPES OF WAVEFORM}
\medskip

The low-frequency gravitational wave discussed in \S3 results
from the Maclaurin-Dedekind transition (region I in Fig.~2).
Apart from this low-frequency wave,
there are two other types of gravitational waveforms that
one may expect, at least in principle, during the evolution
of a Riemann-S ellipsoid due to gravitational radiation.

$\bullet$ {\it ``Spin-up'' Waves\/}:
A Jacobi-like ellipsoid ($|\zeta|<2|\Omega|$)
will evolve to become a stable Maclaurin spheroid if $|\cC|<|\cC_{sec}|$.
If $|\cC|>|\cC_{sec}|$,
it will evolve toward a secularly unstable Maclaurin spheroid, and then,
after a nonaxisymmetric perturbation, it will further evolve
to become a Dedekind ellipsoid. The evolution of
a Jacobi-like ellipsoid toward a Maclaurin spheroid
(regions II and III in Fig.~2) is accompanied by emitting {\it gravitational
wave with increasingly high frequency}.

A special case is the evolution of a Jacobi ellipsoid.
 From Figure 2, we see that a Jacobi ellipsoid will evolve
toward a secularly stable Maclaurin spheroid
\footnote{$^4$}{For a more compressible configuration with $n\go 1.5$,
a Jacobi ellipsoid will first evolve toward a secularly unstable
Maclaurin. This dependence on compressibility comes from the
expansion of the star as it gets more deformed due to rotation
or internal motion. See Fig.~2 in Paper II.}.
Chandrasekhar (1970b) first noted that the rotational frequency increases
as the Jacobi ellipsoid evolves under gravitational radiation. This spin-up is
caused by the decrease of the momentum of inertia as the angular momentum
is radiated away. However, it was assumed that the evolution is along
the equilibrium Jacobi sequence. Thus the evolution ends up at
the bifurcation point. This assumption is not valid, since $\cC$ is
not conserved along a Jacobi sequence. The correct evolutionary
sequence was computed by Miller (1974) for incompressible configurations.

Ipser \& Managan (1984) also considered gravitational wave emission from
the evolution of a compressible Jacobi ellipsoid, based on
their numerical models of rotating polytropes.
They too made the assumption that the evolution is
along a Jacobi sequence, terminating at the bifurcation point.
Because of the mass-shedding constraint, exactly uniform rotating
Jacobi ellipsoid exist only for $n\lo 0.8$. Even for $n=0.5$,
the initial uniformly rotating Jacobi ellipsoid
is very close to the bifurcation point. Thus Ipser \& Managan (1984)
found that only a small fraction ($\sim 10^{-4}-10^{-3}$) of the mass energy
can be radiated as gravitational wave, and the wave is nearly monochromatic
with $f\simeq \Omega_{sec}/\pi\sim 2300-2500$ Hz (see Fig.~5(b)).
This may be somewhat misleading.
Indeed, if slight differential rotation can inhibit the mass shedding
of a Jacobi ellipsoid, as argued in \S 2, a more deformed
triaxial Jacobi configuration might exist in principle. As such a
highly deformed Jacobi ellipsoid evolves toward a Maclaurin spheroid,
much more energy can be emitted, and the frequency can span a wider range.

In Figure 7 we show the gravitational waveforms when a Jacobi-like triaxial
ellipsoid evolves toward a Maclaurin spheroid for $n=1$.
The terminal Maclaurin spheroids (with the same $\cC$ as the initial states)
have $\beta=0.02,~0.12,~0.135$ and $0.24$.
In the first three cases, the evolution ceases once the Maclaurin branch
is reached , while in the last case,
the Maclaurin spheroid (with $\beta=0.24$) can evolve further
to a Dedekind ellipsoid (see \S 3).
In the cases with $\beta=0.12$ and $0.135$, the initial state is chosen to be
a Jacobi ellipsoid. The evolution is
from low-frequency to high (from left to right in Fig.~7).
This is very different from the cases discussed in \S 3.2 (see Fig.~4).
The terminal frequency $f_t$ is determined by the
pattern speed $\Omega_t$ of the ellipsoid as it approaches a Maclaurin shape.
This pattern speed is equal to the angular frequency of the rotating frame
in which the Jacobi mode (with $-$ sign in eq.~[2.4]) becomes neutralized,
i.e., $\sigma_{-}(\Omega_t)=0$. Thus we have
$$\Omega_t={1\over 2}(\Omega_M+\sigma_o),\eqno(4.1)$$
and the terminal (maximum) wave frequency is $f_t=\Omega_t/\pi$.
Figure 8 depicts $f_t$ as a function of $\beta$ of the Maclaurin spheroid.
Note that at the bifurcation point, we have
$\Omega_t=\Omega_M=\Omega_{sec}$, and
at the dynamical stability limit ($\sigma_o=0$), we obtain
$\Omega_t=\Omega_{max}=\Omega_M/2$.

As seen from Figure 7 and Figure 4, the evolution of a Jacobi-like
ellipsoid toward a Maclaurin spheroid is generally much
faster than the evolution of an secularly unstable Maclaurin toward
a Dedekind ellipsoid. This is because a typical ``Jacobi-like''
configuration has larger $|\Omega|$ than a typical ``Dedekind-like''
configuration, and because the energy radiation rate
(eq.~[3.5]) depends on $\Omega$ through a higher-power law.
For the same reasons, the typical wave amplitude during the
``Jacobi-like'' evolution is typically more than ten times larger than
the wave amplitude during the ``Dedekind-like'' evolution.
However, the frequencies of the wave are much larger
for the ``Jacobi-like'' evolution, typically $f\go 1000$ Hz.

The characteristic wave amplitude $h_c$ during the ``Jacobi-like'' evolution
discussed here can also be calculated using equation (3.11). These are also
plotted in Figure 6. Again, $h_c$ can be fitted to the form
$$h_c\simeq 9.1\times 10^{-21}\left({30\,{\rm Mpc}\over D}\right)
M_{1.4}^{3/4}R_{10}^{1/4}f^{-1/5}.\eqno(4.2)$$
This expression is approximately valid for all values of $\cC$ and $n$,
and the differences resulting from different $\cC$, $n$ and initial conditions
lie in the different wave frequency band.
Although $h_c$ during a ``Jacobi-like'' evolution is
larger than $h_c$ during a ``Dedekind-like'' phase, the later
is actually easier to detect since frequency lies around $100$ Hz,
at which LIGO detectors are most sensitive.

$\bullet$ {\it Waves that Carry Negative Angular Momentum\/}:
The third type of gravitational wave results from the evolution
in region IV of Figure 2. The initial configuration
is Dedekind-like ($|\zeta|>2|\Omega|$), but with the pattern speed
$\Omega <0$, i.e., ${\bf \Omega}$ is opposite to the internal vorticity
${\bf \zeta}$ which provides major contribution to the
the angular momentum ${\bf J}$ of the system.
Such a configuration will evolve toward a Dedekind ellipsoid
(with terminal $\Omega=0$) when $|\cC|>|\cC_{sec}|$,
or toward a stable Maclaurin spheroid
(with terminal $\Omega<0$; see eq.~[3.9]) when $|\cC|<|\cC_{sec}|$.
Figure~9 depicts two examples of such evolution.
Note that $\Omega$ does not change monotonically as the
$a_2/a_1$ decreases. For an extremely large deformation (small $a_2/a_1$),
${\bf J}$ is in the same direction as ${\bf \Omega}$, and
$|J|$ decreases and $|\Omega|$ increases during the evolution.
When the deformation is not too large, the total
angular momentum of the system
is positive due to the large positive internal vorticity $\zeta$.
As the system evolves toward the terminal state (Dedekind or
Maclaurin), {\it the energy decreases but the angular momentum $|J|$
increases, while the ellipsoid spins down} ({\it $|\Omega|$ decreases}).
This peculiar behavior comes about because $\Omega$ is negative,
and $dE=\Omega dJ$ must be satisfied for secular evolution
via gravitational radiation dissipation (Ostriker \& Gunn 1969).
The gravitational wave carries away negative angular momentum,
resulting in an increase of the angular momentum of the system.
The wave frequency can also be low, but the wave amplitude
decreases monotonically.

It is not clear whether the waveforms discussed in this section
is ever relevant to a new-born neutron star. This depends on whether
the collapsed core is axisymmetric or nonaxisymmetric after hydrostatic
equilibrium is established following the initial
implosion and/or the dynamical instability.
Earlier calculations by Durisen et al.~(1986),
Williams \& Tohline (1988) found that a dynamically unstable
rotating star evolves toward a triaxial bar surrounded by a ring.
Recent calculations by Houser et al (1994), which include shock dissipation
of the ejected gas, suggest that the final
outcome is an axisymmetric core surround by a shocked halo.
Clearly this issue will be resolved only by
three-dimensional collapsing simulations, which have yet to be performed
\footnote{$^5$}{In the binary coalescence calculation
of Rasio \& Shapiro (1994), a Jacobi-like triaxial system
is produced for a sufficiently small polytropic index.
However, this calculation starts from a corotating binary near contact.
Calculations starting from non-synchronized spins (more realistic
for neutron stars)  may yield a different result;
see Shibata, Nakamura \& Oohara 1992.}.
Nevertheless, the gravitational waveforms discussed in \S 3.3
and in the present section represent the three classes
of waveforms that can be expected from the secular non-axisymmetric
evolution of any rotating star.

\bigskip
\centerline{\bf 5. EFFECTS OF VISCOSITY}
\medskip

So far in our discussion we have assumed that the
star consists of inviscid fluid. When the fluid viscosity is
sufficiently large, it tends to counteract the GR instability (Lindblom \&
Detweiler 1977). From analysis in Appendix A, the combined effect
of viscosity and GR reaction is determined by the ratio of the gravitational
radiation timescale and the viscous timescale, $Q\sim t_{GW}/t_{visc}$,
or, more precisely (eq.~[A33])
$$Q=4.4\times 10^{-13}\left(1-{n\over 5}\right)^2
R_{10}^2M_{1.4}^{-3}{\bar\nu},\eqno(5.1)$$
where ${\bar\nu}$ is the mean kinematic shear viscosity in units of cm$^2/$s.
Figure 10 shows the critical $T/|W|$ for instability as a function of $Q$
for a Maclaurin spheroid with $n=0,~0.5,~1$ and $1.5$.
For $Q\lo 1$, the instability is mainly caused by GR reaction in the
``Dedekind-like'' mode, while for $Q\go 1$,
it is mainly induced by viscous stress in the ``Jacobi-like'' mode.
In the limits of $Q>>1$ and $Q<<1$, both instabilities set in at the
bifurcation
point $\beta_{sec}=0.1375$. For $Q\sim 1$, both modes are stablized, and
the Maclaurin spheroid can be stable all the way to the dynamical
limit for certain value of $Q\sim 1$.

The secular evolution discussed in \S2-4 typically occurs
during the first $1-100$ seconds after the neutron star's birth.
During this epoch, the neutron star remains very hot, with temperature
$T\go 10^{10}$ K (Burrows \& Lattimer 1986).
The shear viscosity resulting from neutron-neutron scattering
is ${\bar\nu}_n\sim 10 \rho_{15}^{5/4}T_{10}^{-2}$ cm$^2/$s (Flowers \& Itoh
1976, with the fitting formula from Cutler \& Lindblom 1987),
where $\rho_{15}$ is the density in units of $10^{15}$ g/cm$^3$, and
$T_{10}=T/(10^{10}~{\rm K})$. This is much too small to have any effect
on the secular instability and evolution discussed in \S 2-4.
The neutrino-induced shear viscosity (resulting from neutrino-neutron
scattering) ${\bar\nu}_\nu \sim 10^6\rho_{15}^{-4/3}T_{10}$ cm$^2/$s
(Thompson \& Duncan 1993) is also too small to be of any importance.

There have been some discussions about
the importance of neutron star bulk viscosity
in the literature. For a hot neutron star, a bulk viscosity can arise
from the phase lag between density and pressure
perturbation due to the relatively long time scale of weak interactions to
re-establish chemical equilibrium (Saywer 1989).
In our ellipsoid model, the bar-mode is associated
with the perturbations of the three axes given by
$\delta a_1=-\delta a_2$, $\delta a_3=0$. Therefore
the bulk viscosity has no effect on the damping/growth time of the mode.
In realistic calculations, however, there is a small but nonzero
compression/expansion of the fluid associated with the bar mode,
thus bulk viscosity can play a role (Ipser \& Lindblom 1991).
Using the Sawyer's formula for bulk viscosity ${\bar \nu_B}\sim
10^{11}\rho_{15}T_{10}^6$ cm$^2/$s (where we have taken the
perturbation timescale to be of order a millisecond), Ipser \& Lindblom (1991)
found that the bulk viscosity can suppress the GR driven instability
for $T\go 2\times 10^{10}$ K
\footnote{$^6$}{Ipser \& Lindblom only considered uniformly
rotating models and were thus restricted to rotation frequencies
below the mass shedding limit.}.
However, it should be noted
that the above formula for bulk viscosity assumes that the neutrinos
produced in electron-capture escape the neutron star freely. Thus
the formula is valid only for relatively low temperatures
(less than a few times $10^9$ K; e.g., Shapiro \& Teukolsky 1983).
During the early epoch of the neutron star studied in this paper,
the neutrino optical depth is large,
and the bulk viscosity is likely to be greatly suppressed due the
blocking of the
neutrino phase space. 
A more detailed study is needed to fully address this issue.

In the presence of a shear viscosity, the fluid circulation evolves
according to
$${d\cC\over dt}=-\bar\nu M\Lambda\left({a_1^2-a_2^2\over a_1a_2}\right)^2,
\eqno(5.2)$$
(cf.~Paper II). This result and equation (3.5) for $dJ/dt$ govern the secular
evolution of a general ellipsoid under the combined effects
of viscosity and GR reaction. Some evolutionary tracks have been
considered by Detweiler \& Lindblom (1977) for incompressible
configurations.

Now let us consider the effect of a small viscosity on the secular evolution.
Without viscosity, we see from \S 3 that a Dedekind ellipsoid is one of the
possible final states of the evolution. When viscosity is also present,
however small, only a secularly stable Maclaurin spheroid can be the
final state --- it is the only configuration which does not
radiate GR or dissipate energy viscously.
As the star approaches a Dedekind ellipsoid,
the gravitational evolution timescale increases.
When this timescale becomes
comparable to the viscous dissipation timescale, the star will be
driven along a nearly-Dedekind sequence, eventually toward the
bifurcation point. Thus we can expect that
{\it many rapidly rotating nascent neutron stars
eventually settle down at the bifurcation point along the Maclaurin
sequence\/}. Of course, the stars can then be subjected to further spin-down
due to magnetic braking.

\bigskip
\centerline{\bf 6. DISCUSSIONS}
\medskip

Thorne (1987,~1992) has summarized the three radically different
scenarios of rapidly rotating collapse and the resulting wave
characteristics: (i) the star remains axisymmetric
throughout the collapse, resulting in a weak wave signal;
(ii) the star becomes a (Jacobi-like) triaxial object, emitting nearly
monochromatic waves with $f\sim 1000$ Hz;
(iii) the core breaks up into lumps, resulting in the strongest wave signal.
Regarding (ii), we have shown in \S 4 that in general the wave emitted
by a Jacobi-like rotating star can span a wide frequency band up to a few
thousands Hertz. Moreover, we have considered in this paper
the gravitational wave signal in another
likely scenario for the post-collapse, quasi-static stage:
After the hydrodynamical collapse which lasts $\sim 1$ second,
the core settles down into a nearly axisymmetric, equilibrium
state, which is secularly unstable to nonaxisymmetric
perturbations. Prior to this, the star may have gone through a
nonaxisymmetric dynamical phase, but the development of
the dynamical instability can only render the final configuration dynamically
stable and not necessarily secularly stable (Houser et al 1994).
Due to the bar-mode instability of gravitational radiation reaction
(Chandrasekhar 1970a, Friedman \& Schutz 1978), the nascent neutron star
evolves quasi-statically toward a
Dedekind-like object on a timescale of tens of seconds to a few minutes,
emitting gravitational waves with frequency
sweeping from a few hundred Hertz down to zero, and a non-monotonic
wave amplitude. The viscous forces finally bring the
Dedekind-like neutron star to an axisymmetric Maclaurin spheroid near the
bifurcation point on a much longer, viscous timescale.

Gravitational radiation resulting from the
Chandrasekhar-Friedman-Schutz instability has been discussed
in the context of accreting neutron stars (Wagoner 1984). As a slowly
rotating neutron star accretes mass from a binary companion,
it can be spun up until it becomes rotationally unstable on
an accretion timescale. A monochromatic gravitational wave
is then radiated, carrying off the angular momentum of
the accreted material. During accretion, which increases the star's angular
momentum very slowly, the neutron star can evolve only slightly above the
stability limit. Accordingly it is often the instability of a higher-order
($m=4-6$) mode (which sets in at smaller $\beta$) that is relevant,
and the star can never reach the
bar-mode ($m=2$) stability limit (at $\beta=\beta_{sec}$).
Much effort has been devoted in calculating accurately
the stability limits of these high-order modes
(e.g., Ipser \& Lindblom 1990,~1991).
By contrast, in the core collapse considered in this paper,
the nascent neutron star can easily reach a value of $\beta$ much larger than
the secular stability limit $\beta_{sec}$, as a result of the much violent
initial implosion. Thus in this case, the dominant instability
is the bar-mode. Schutz (1989) has also noted that the instability-driven
spin-down can yield low frequency gravitational waves,
but he has focused on the high-order modes. As we noted before,
unlike the bar-mode instability, the quantitative outcome
of a high-order instability is not yet clear at present.

One important question we have not addressed so far is the
size $R_o$ of the collapsed core. The gravitational wave frequency
depends sensitively on this quantity through $f\propto R_o^{-3/2}$
(cf.~eq.~[3.10]). We have adopted $R_o=10$ km throughout the paper.
While the thermal pressure
is unlikely to affect the radius of the collapsed core appreciably
(\S 2), it is possible that the collapse can be stopped
by centrifugal forces at sub-nuclear density.
Such ``centrifugal'' bounce has been found in some
numerical simulations of axisymmetric collapse
(M\"onchmeyer et al.~1991, Yamada \& Sato 1993).
If this indeed occurs, the wave frequency would be lower and the
evolution timescale longer than discussed in this paper,
although the characteristics of the waves do not change.
On the other hand, it is conceivable that any realistic rotating core will
ultimately contract to typical neutron star size. This must be the case
if we believe that most neutron stars are born with high spin rate.
Actual detection of gravitational waves from newly-formed neutron stars
might provide valuable information about the
size and rotational properties of the collapsed cores and the dynamics
of core collapse itself.

\bigskip
\bigskip

This work has been supported in part
by NSF Grant AST 91-19475 and NASA Grant NAGW-2364 to Cornell University.

\bigskip
\bigskip
\bigskip
\centerline{\bf APPENDIX A: THE SECULAR INSTABILITY GROWTH TIMES OF}
\smallskip
\centerline{\bf COMPRESSIBLE MACLAURIN SPHEROIDS}
\medskip

To calculate the oscillation modes of a Maclaurin spheroid,
we start from the general dynamical equations for a Riemann-S
ellipsoid, including the viscous forces and gravitational radiation (GR)
reaction (see Paper II, \S 2,4):
\footnote{$^7$}{A bulk viscosity contribution, which was not included in Paper
II, can be easily incorporated
by adding a term $-(5{\bar\nu_B}/\kappa_n)(\dot a_1/a_1+\dot a_2/a_2
+\dot a_3/a_3)/a_1$ to the right hand side of eq.~(A1) and similar terms
to eqs.~(A2)-(A3), where $\bar\nu_B$ is the mean kinematic bulk viscosity.
However, this will have no effect on the
bar-mode in which we are most interested here.}
$$\eqalignno{
&\ddot a_1 = a_1(\Omega^2+\Lambda^2)-2a_2\Omega\Lambda
-{2\pi\over q_n}a_1A_1\bar\rho
+\left({5k_1 \over n\kappa_n}\right){P_c\over\rho_c}{1\over a_1} \cr
&~~~~~~~~~~~~~~~~~~~~~~~~~~~~~~~~~~ -{10\over 3\kappa_n}\bar\nu\left(
	{2\dot a_1\over a_1}-{\dot a_2\over a_2}-{\dot a_3\over a_3}\right)
	{1\over a_1}-{2\over 5}\Ib5_{11}a_1,  &(A1)\cr
&\ddot a_2 = a_2(\Omega^2+\Lambda^2)-2a_1\Omega\Lambda
-{2\pi\over q_n}a_2A_2\bar\rho
+\left({5k_1 \over n\kappa_n}\right){P_c\over\rho_c}{1\over a_2}\cr
&~~~~~~~~~~~~~~~~~~~~~~~~~~~~~~~~~~ -{10\over 3\kappa_n}\bar\nu\left(
	{2\dot a_2\over a_2}-{\dot a_1\over a_1}-{\dot a_3\over a_3}\right)
	{1\over a_2}-{2\over 5}\Ib5_{22}a_2,  &(A2)\cr
&\ddot a_3 =-{2\pi\over q_n}a_3A_3\bar\rho
+\left({5k_1 \over n\kappa_n}\right){P_c\over\rho_c}{1\over a_3}
-{10\over 3\kappa_n}\bar\nu\left(
{2\dot a_3\over a_3}-{\dot a_1\over a_1}-{\dot a_2\over a_2}\right)
{1\over a_3}-{2\over 5}\Ib5_{33}a_3,  &(A3)\cr
&{d\over dt}\left(a_1\Omega-a_2\Lambda\right)
=-\dot a_1\Omega+\dot a_2\Lambda
-{5\over\kappa_n}\bar\nu {a_1^2-a_2^2\over a_1^2a_2}\Lambda
-{2\over 5}\Ib5_{12}a_1,       &(A4)\cr
&{d\over dt}\left(-a_2\Omega+a_1\Lambda\right)
=\dot a_2\Omega-\dot a_1\Lambda
-{5\over\kappa_n}\bar\nu {a_1^2-a_2^2\over a_1a_2^2}\Lambda
-{2\over 5}\Ib5_{12}a_2,       &(A5)\cr
}$$
The dynamical varibles
are the three axes $a_i$ ($i=1,2,3$), the angular frequency of the figure
(pattern speed) $\Omega$, and the angular frequency $\Lambda$ specifying
the internal fluid motion (relative to the figure) with uniform
vorticity $\zeta=-[(a_1^2+a_2^2)/a_1a_2]\Lambda$.
The other quantities have the same meaning as in Paper II.

When linearizing the dynamical equations,
we must allow for the fact that a Maclaurin spheroid need not
be assigned the angular frequency of the frame in which it is at rest.
When viewed from a rotating frame with angular frequency $\Omega$ relative to
the inerial frame, the Maclaurin spheroid will appear as having stationary
internal motion with the uniform vorticity $\zeta=2(\Omega_M-\Omega)$,
i.e., $\Lambda=\Omega-\Omega_M$, where $\Omega_M$ is the angular frequency
of the Maclaurin spheroid relative to an inertial frame.
Thus the values of $\Omega_{eq}$ and $\Lambda_{eq}$
for the equilibrium Maclaurin spheroid can in principle
be arbitrary, subject only to the condition
$$\Omega_{eq}-\Lambda_{eq}=\Omega_M.\eqno(A6)$$

Consider small perturbations of the form
$$a_i=a_{i,eq}(1+\alpha_i),~~~\Omega=\Omega_{eq}+\omega,~~~
\Lambda=\Lambda_{eq}+\lambda,\eqno(A7)$$
with $\alpha_i\propto e^{-i\sigma t}$ and similarly for $\omega$, $\lambda$.
Note that $\Omega_{eq}$ (or $\Lambda_{eq}$) must be determined from the
linear mode analysis. We are only interested in the non-axisymmetric
bar-mode (called ``toroidal mode'' in Ch69) for which it turns out
$\alpha_1=-\alpha_2$ and $\alpha_3=0$ (see Paper II, \S3.3).

\bigskip
\centerline{\bf A.1 Pure Dynamical Oscillations}
\medskip

First consider the case when viscosity and GR reaction are
absent. A useful expression is
$${a_j\over A_i}{\partial A_i\over\partial a_j}
=\cases{-2+{3B_{ii}/A_i}, & if $i=j$;\cr
	{B_{ij}/A_i},      & if $i\neq j$,\cr}
\eqno(A8)$$
(no summation for repeated indices), where $A_i$, $B_{ij}$ are defined as
in Ch69 (Chap.~3). Also it is useful to
note that the pressure term in (A1)-(A3) can be written
as $5k_1P_c/(n\kappa_n\rho_c)=(M/q_nR_o)(R_o/R)^{3/n}=5U/(n\kappa_nM)$,
where $U$ is the internal energy (Paper II, eq.~[2.9]).
Linearizing equation (A1) and using
the equilibrium conditions (Paper II, eq.~[3.24]), we obtain
$$\eqalign{
I_{11}\ddot\alpha_1 =& I_{11}\biggl[\Omega_M^2\alpha_1+2\Omega_{eq}
\Lambda_{eq} (\alpha_1-\alpha_2)+2\Omega_M(\omega-\lambda)\biggr]\cr
& +\Sigma a_1^2\biggl[(3B_{11}-2A_1)\alpha_1+(B_{11}-A_1)\alpha_2
+(B_{13}-A_1)\alpha_3\biggr]\cr
& -\Sigma a_3^2A_3\biggl[-\alpha_1-
(\alpha_1+\alpha_2+\alpha_3)/n\biggr],\cr
}\eqno(A9)$$
where the notation is the same as in Paper II (\S 3.3), with
$I_{11}=\kappa_n Ma_1^2/5$ and $\Sigma\equiv -{3M^2/[2(5-n)R^3]}$.
The equation for $\ddot\alpha_2$ can be obtained by inter-changing
subscript $1$ and $2$ in (A9).
Linearizing equations (A4)-(A5) yields
$$\Omega_{eq}=-\Lambda_{eq}={\Omega_M\over 2},\eqno(A10)$$
and
$$\omega-\lambda=-\Omega_M (\alpha_1+\alpha_2), \eqno(A11)$$
where we have used $\Omega_{eq}-\Lambda_{eq}=\Omega_M$ (eq.~[A6]).
Substituting (A10)-(A11) into (A9) and the corresponding equation
for $\ddot\alpha_2$, we can see that $\alpha_1=-\alpha_2$, $\alpha_3=0$ is
indeed a solution. After some algebra, we obtain
$$I_{11}\Delta\ddot\alpha=-I_{11}\sigma_o^2\Delta\alpha,\eqno(A12)$$
where $\Delta\alpha\equiv\alpha_1-\alpha_2$, and
$$\sigma_o=\left({4\over q_n}\pi\bar\rho
B_{11}-\Omega_M^2\right)^{1/2}.\eqno(A13)$$
Thus the mode frequency is $\sigma=\pm\sigma_o$, in agreement with the result
of Paper II, obtained using a Hamiltonian formalism.
Note that the frequency $\sigma_o$ is the viewed in a particular
rotating frame with $\Omega=\Omega_M/2$, in which the oscillation is
self-adjointed ($\Omega=-\Lambda$), and the directions of the
principal axes of the ellipsoid are fixed (Ch69, \S 36).
In a frame with a different
$\Omega$, the oscillation frequency is
$$\sigma_\pm(\Omega)=\pm\sigma_o+2\left(\Omega-{\Omega_M\over
2}\right),\eqno(A14)$$
since this mode corresponds to a $l=m=2$ mode with a $\exp(i2\phi)$ dependence
on the azimuthal angle (see Ch69, \S36; Rossner 1967).

\bigskip
\centerline{\bf A.2 Effect of Viscosity}
\medskip

Now consider a small viscosity $\bar\nu$
so that we only consider the leading correction $\sim {\cal O}(\bar\nu)$
to the mode frequency.
Again assuming a $e^{-i\sigma t}$ time dependence of the
perturbations, from equations (A4)-(A6) we obtain
l$$\Omega_{eq}={1\over 2}\Omega_M\left(1+i{5\bar\nu\over\kappa_na_1^2\sigma}
\right),~~~\Lambda_{eq}=-{1\over 2}\Omega_M\left(1-
i{5\bar\nu\over\kappa_na_1^2\sigma}\right),\eqno(A15)$$
while equation (A11) still holds.
The fact that $\Omega_{eq}$ contains an imaginary part implies that
the rotation of the frame in which the oscillation is specified
can itself enhance or damp the oscillation amplitude.
Substituting (A11) and (A15) into the linearized equations
for $\ddot \alpha_i$, we obtain
$$\Delta\ddot\alpha=-\sigma_o^2\Delta\alpha
-{10\bar\nu\over\kappa_n a_1^2}\Delta\dot\alpha,\eqno(A16)$$
from which we have
$$\sigma=\pm\sigma_o-i{5\bar\nu\over\kappa_na_1^2}.
\eqno(A17)$$
However, this is the frequency of oscillation in a frame
with angular frequency given by a complex $\Omega_{eq}$.
Similar to equation (A14), in a general frame with (real) angular velocity
$\Omega$, the oscillation frequency is
$$\eqalign{
\sigma_\pm(\Omega) &=\pm\sigma_o-i{5\bar\nu\over\kappa_na_1^2}
+2\left[\Omega-\left({\Omega_M\over 2}\pm i{5\bar\nu\Omega_M
\over 2\kappa_n a_1^2\sigma_o}\right)\right]	\cr
&=\pm\sigma_o+2\left(\Omega-{\Omega_M\over 2}\right)
-i{5\bar\nu\over\kappa_na_1^2}{\sigma_o\pm\Omega_M\over\sigma_o}.\cr
}\eqno(A18)$$
Clearly, the Dedekind-mode (with upper sign in eq.~[A18])
is always secularly stable. For the Jacobi-mode (with lower sign in
eq.~[A18]), secular instability sets in when $\Omega_M>\sigma_o$, or when
$$q_n{\Omega_M^2\over\pi\bar\rho}>2B_{11},
\eqno(A19)$$
corresponding to bifurcation values $e=0.8127$ and $T/|W|=0.1375$.
The instability growth time $\tau_{vis}$ is given by
$$\tau_{vis}^{-1}={5\bar\nu\over\kappa_na_1^2}
\left({\Omega_M-\sigma_o\over\sigma_o}\right).
\eqno(A20)$$

\bigskip
\centerline{\bf A.3 Effect of Gravitational Radiation}
\medskip

Now to include the GR reaction, we first need to calculate
$\Ib5_{ij}$, the fifth time derivative of the reduced quadrupole moment
tensor of the body in the inertial frame projected onto the
rotating body frame (with basis vectors along the principal axes).
This is done using the procedure given in Miller (1974) and
Appendix A of Paper II. Unlike the quasi-static case (as in
Appendix A of Paper II), in order to discern the instability,
$\Ib5_{ij}$ need to be calculated to all order.
For small perturbations around a Maclaurin spheroid, we obtain
$$\eqalign{
\biggl[\Ib5_{ij}\biggr]= &
\left[15\Omega^5(I_{11}-I_{22})-40\Omega^3(I_{11}^{(2)}-I_{22}^{(2)})
+5\Omega(I_{11}^{(4)}-I_{22}^{(4)})\right]
\left(\matrix{0 & 1 & 0\cr
              1 & 0 & 0\cr
              0 & 0 & 0\cr}\right)\cr
&+\left[40\Omega^4(I_{11}^{(1)}-I_{22}^{(1)})-20\Omega^2(I_{11}^{(3)}
-I_{22}^{(3)})\right]
\left(\matrix{1 & 0 & 0\cr
              0 & -1& 0\cr
              0 & 0 & 0\cr}\right)\cr
&+\left(\matrix{\Ib5_{11} & 0 & 0\cr
              0 & \Ib5_{22}& 0\cr
              0 & 0 & \Ib5_{33}\cr}\right),\cr
}\eqno(A21)$$
where on the right-hand side, the superscript $(i)$ now means $i$-th time
derivative in the frame of the body. Thus
the only non-trivial components are therefore
$$\eqalign{
\Ib5_{11}=&-{40\over 5}i\kappa_nMa_1^2(\alpha_1-\alpha_2)(2\Omega^4\sigma
+\Omega^2\sigma^3)-{2\over 5}i\kappa_nMa_1^2\sigma^5\alpha_1
-{1\over 3}I_t^{(5)},\cr
\Ib5_{22}=&+{40\over 5}i\kappa_nMa_1^2(\alpha_1-\alpha_2)(2\Omega^4\sigma
+\Omega^2\sigma^3)-{2\over 5}i\kappa_nMa_1^2\sigma^5\alpha_2
-{1\over 3}I_t^{(5)},\cr
\Ib5_{33}=&-{2\over 5}i\kappa_nMa_3^2\sigma^5\alpha_3
-{1\over 3}I_t^{(5)},\cr
\Ib5_{12}=&\Ib5_{21}={2\over 5}\kappa_nMa_1^2(\alpha_1-\alpha_2)
(16\Omega^5+40\Omega^3\sigma^2+5\Omega\sigma^4),\cr
}\eqno(A22)$$
where $I_t\equiv I_{11}+I_{22}+I_{33}=\kappa_nM(a_1^2+a_2^2+a_3^2)/5$.

Linearizing equations (A4)-(A5) including GR reaction alone, we have
$$\eqalign{
\Omega_{eq}=&{1\over 2}\Omega_M
-i{2\kappa_nMa_1^2\over 25}{1\over\sigma}(16\Omega_{eq}^5+
40\Omega_{eq}^3\sigma^2+5\Omega_{eq}\sigma^4)\cr
\simeq&{1\over 2}\Omega_M
-i{\kappa_nMa_1^2\over 25}{1\over\sigma}(\Omega_M^5+
10\Omega_M^3\sigma^2+5\Omega_M\sigma^4)\cr
}\eqno(A23)$$
and $\Lambda_{eq}=\Omega_{eq}-\Omega_M$, where we have
assumed that the GR reaction is a small correction.
Linearizing equations for $\ddot a_i$, we obtain, for the bar-mode
$$\eqalign{
\Delta\ddot\alpha=&-\sigma_o^2\Delta\alpha-
{2\over 5}\left(\Ib5_{11}-\Ib5_{22}\right)\cr
=&-\sigma_o^2\Delta\alpha+i{4\kappa_nMa_1^2\over 25}
\left[40(2\Omega_{eq}^4\sigma+\Omega_{eq}^2\sigma^3)+\sigma^5\right]
\Delta\alpha.\cr
}\eqno(A24)$$
With equation (A23), we then have
$$\sigma=\pm\sigma_o-i{2\kappa_nMa_1^2\over 25}
\left[5(\Omega_M^4+2\Omega_M^2\sigma_o^2)+\sigma_o^4\right].
\eqno(A25)$$
Again, this is only the frequency of oscillation in a frame
with the complex angular frequency $\Omega_{eq}$. The existence of an imaginary
part in $\Omega_{eq}$ implies additional mode damping or growing.
Similar to (A14), in a general frame with (real) angular velocity
$\Omega$, the oscillation frequency is
$$\eqalign{
\sigma_\pm(\Omega) &=\pm\sigma_o-i{2\kappa_nMa_1^2\over 25}
\left[5(\Omega_M^4+2\Omega_M^2\sigma_o^2)+\sigma_o^4\right]
+2\left(\Omega-\Omega_{eq}\right)\cr
&=\pm\sigma_o+2\left(\Omega-{\Omega_M\over 2}\right)
-i{2\kappa_nMa_1^2\over 25}{(\sigma_o\mp\Omega_M)^5\over\sigma_o}.
}\eqno(A26)$$
Clearly, the Dedekind-mode become secularly unstable when
equation (A19) is satisfied, i.e., the gravitational radiation also
induces an instability at the bifurcation point, but for a different mode
as for viscosity. The growth time $\tau_{GW}$ is given by
$$\tau_{GW}^{-1}={2\kappa_nMa_1^2\over 25}
{(\Omega_M-\sigma_o)^5\over\sigma_o}.\eqno(A27)$$

\bigskip
\centerline{\bf A.4 Combined Effects of Viscosity and Gravitational Radiation}
\medskip

The secular bar-mode instability tends to weaken or diminish
when both viscosity and GR reaction are present.
This is because viscous dissipation and GR reaction cause
different modes to become unstable, as seen from \S A.2 and \S A.3.
The combined effects of viscosity and GR reaction
on the secular instability of incompressible Maclaurin spheroid
were first noted by Lindblom \& Detweiler (1977).
Combining the results of \S A.2 and \S A.3, we find that
the (complex) frequency for the bar mode in a rotating frame with
angular velocity $\Omega$is given by:
$$\sigma_\pm(\Omega)=\pm\sigma_o+(2\Omega-\Omega_M)
-i\left[{5\bar\nu\over\kappa_na_1^2}{\sigma_o\pm\Omega_M\over\sigma_o}
+{2\kappa_nMa_1^2\over 25}{(\sigma_o\mp\Omega_M)^5\over\sigma_o}\right].
\eqno(A28)$$
For convenience, let us
define dimensional viscous time constant and gravitational radiation
time constant via
$$t_{vis}^{-1}={5\bar\nu\over\kappa_nR_o^2},~~~~~
t_{GW}^{-1}={9\kappa_n\over 200 q_n^2}\left({GM\over R_oc^2}\right)^{5/2}
\left({GM\over R_o^3}\right)^{1/2},\eqno(A29)$$
($G,~c$ have be restored).
The growth time $\tau_\pm>0$ of the two modes are then given by
$$\tau_{+}^{-1}=t_{GW}^{-1}
\left[(1-e^2)^{-1/3}\left({R_o\over R}\right)^4{(\hat\Omega_M-\hat\sigma_o)^5
\over\hat\sigma_o}\right]
-t_{vis}^{-1}\left[(1-e^2)^{1/3}\left({R_o\over R}\right)^2
{\hat\Omega_M+\hat\sigma_o\over\hat\sigma_o}\right],
\eqno(A30)$$
and
$$\tau_{-}^{-1}=t_{vis}^{-1}\left[(1-e^2)^{1/3}\left({R_o\over R}\right)^2
{\hat\Omega_M-\hat\sigma_o\over\hat\sigma_o}\right]
-t_{GW}^{-1}
\left[(1-e^2)^{-1/3}\left({R_o\over R}\right)^4{(\hat\Omega_M+\hat\sigma_o)^5
\over\hat\sigma_o}\right],
\eqno(A31)$$
where the quantities in $[\cdots]$ are dimensionless and depend only on
the eccentricity $e$ (for compressible systems, there is also a dependence
on $n$ through $R_o/R$), and
$$\hat\Omega_M={\Omega_M\over\sqrt{\pi\bar\rho}}q_n^{1/2},~~~
\hat\sigma_o={\sigma_o\over\sqrt{\pi\bar\rho}}q_n^{1/2}
=(4B_{11}-\hat\Omega_M^2)^{1/2}.
\eqno(A32)$$
When $\tau_\pm<0$, the mode involved is stable, and $|\tau_\pm|$
then gives the corresponding damping time.
In the limit of $n=0$, the above results agree with that of
Lindblom and Detweiler (1977).

It is clear from the above expressions that the bar mode of
a a compressible Maclaurin spheroid
tends to be stablized by the combined effects of viscosity and GR reaction.
To locate the region of stability, define
$$Q\equiv {t_{GW}\over t_{vis}}
(1-e_d^2)^{2/3}\hat\Omega_d^{-4},\eqno(A33)$$
where $e_d=0.95289$ and $\hat\Omega_d^2=q_n\Omega_d^2/(\pi\bar\rho)
=0.44022$ refer to the dynamical stability limit.
The Dedekind-mode is stable when
$$Q\hat\Omega_d^4\left({1-e^2\over 1-e_d^2}\right)^{2/3}
\left({R\over R_o}\right)^2
(\hat\Omega_M+\hat\sigma_o)\ge (\hat\Omega_M-\hat\sigma_o)^5,\eqno(A34)$$
and the Jacobi-mode is stable when
$$Q\hat\Omega_d^4\left({1-e^2\over 1-e_d^2}\right)^{2/3}
\left({R\over R_o}\right)^2
(\hat\Omega_M-\hat\sigma_o)\le (\hat\Omega_M+\hat\sigma_o)^5.\eqno(A35)$$
Figure 9 depicts the critical $T/|W|$ (where the
secular instability first sets in) as a function of $Q$ for
$n=0,~0.5,~1,~1.5$. For $n=0$ and $Q=1$, the Maclaurin spheroid
is stable all the way to the dynamical limit.

\vfill\eject
\centerline{\bf REFERENCES}
\medskip
\def\bysame{\hbox to 50pt{\leaders\hrule height 2.4pt depth -2pt\hfill .\ }}
\def\hi{\noindent \hangindent=2.5em}

\hi{
Abramovici, A., et al. 1992, Science, 256, 325}

\hi{
Bahcall, J.A., \& Piran, T. 1983, ApJ, 267, L77}

\hi{
Bailyn, C.D., \& Grindlay, J.E. 1990, ApJ, 353, 159}

\hi{
Blair, D. 1989, in {\it Gravitational Wave Data Analysis\/}, ed. B.F.~Schutz
(Dordrecht: Kluwer)}

\hi{
Bodenheimer, P., \& Ostriker, J.P. 1973, ApJ, 180, 159}

\hi{
Bradaschia, C., et al. 1990, Nucl. Instrum. \& Methods, A289, 518}

\hi{
Burrows, A. 1990, in {\it Supernovae\/}, ed. A.G.~Petschek
(Springer-Verlag: New York)}

\hi{
Burrows, A., \& Lattimer, J.M. 1986, ApJ, 307, 178}

\hi{
Chandrasekhar, S. 1969, Ellipsoidal Figures of Equilibrium
(New Haven: Yale University Press) (Ch69)}

\hi{
\bysame 1970a, ApJ, 161, 561}

\hi{
\bysame 1970b, ApJ, 161, 571}

\hi{
Comins, N. 1979a, MNRAS, 189, 233}

\hi{
\bysame 1979b, MNRAS, 189, 255}

\hi{
Cordes, J.M., Romani, R.W., \& Lundgren, S.C. 1993, Nature, 362, 133}

\hi{
Cropper, M., et al. 1988, MNRAS, 231, 695}

\hi{
Cutler, C., \& Lindblom, L. 1987, ApJ, 314, 234}

\hi{
Detweiler, S.L., \& Lindblom, L. 1977, ApJ, 213, 193}

\hi{
Durisen, R.H., Gingold, R.A., Tohline, J.E., \& Boss, A.P. 1986, ApJ, 305, 281}


\hi{
Finn, L.S. 1991, in {\it Nonlinear Problems in Relativity and Cosmology\/}
(Annals of the New York Academy of Sciences, Vol.~631)
ed. J.R.~Buchler, S.L.~Detweiler \& J.R.~Ipser (New York)}


\hi{
Flowers, E., \& Itoh, N. 1976, ApJ, 206, 218}

\hi{
Friedman, J., \& Schutz, B.F. 1978, ApJ, 222, 281}

\hi{
Houser, J.L., Centrella, J.M., \& Smith, S.C. 1994, Phys. Rev. Lett., 72, 1314}

\hi{
Imamura, J.N., Friedman, J.L., \& Durisen, R.H. 1985, ApJ, 294, 474}

\hi{
Ipser, J.M., \& Lindblom 1990, ApJ, 355, 226}

\hi{
\bysame 1991, ApJ, 373, 213}

\hi{
Ipser, J.R., \& Managan, R.A. 1984, ApJ, 282, 287}

\hi{
James, R.A. 1964, ApJ, 140, 552}


\hi{
Lai, D., Rasio, F.A., \& Shapiro, S.L. 1993, ApJS, 88, 205 (Paper I)}




\hi{
\bysame 1994, ApJ, submitted (Paper II)}

\hi{
Landau, L.D., \& Lifshitz, E.M. 1987, Fluid Mechanics, 2nd
Ed.\ (Oxford: Pergamon Press)}

\hi{
Lindblom, L, \& Detweiler 1977, ApJ, 211, 565}

\hi{
Managan, R.A. 1985, ApJ, 294, 463}

\hi{
Michel, F.C., 1987, Nat., 329, 310}


\hi{
Miller, B.D. 1974, ApJ, 187, 609}

\hi{
Misner, C.M., Thorne, K.S., \& Wheeler, J.A. 1970,
{\it Gravitation\/} (New York: Freeman)}

\hi{
M\"onchmeyer, R., Sch\"afer, G., M\"uller, E., \& Kates, R.E. 1991,
A\&A, 246, 417}

\hi{
Ostriker, J.P., \& Bodenheimer, P. 1973, ApJ, 180, 171}

\hi{
Ostriker, J.P., \& Gunn, J.E. 1969, ApJ, 157, 1395}

\hi{
Radhakrishnan, V. 1992, in {\it X-ray Binaries and Recycled Pulsars},
ed.~E.P.J. van den Heuvel and S.A. Rappaport (Kluwer: Netherlands)}

\hi{
Rasio, F.A., \& Shapiro, S.L. 1994, ApJ, in press}

\hi{
Roberts, P.H., \& Stewartson, K. 1963, ApJ, 137, 777}

\hi{
Rossner, L.F. 1967, ApJ, 149, 145}

\hi{
Saenz, R.A., \& Shapiro, S.L. 1981, ApJ, 244, 1033}

\hi{
Sawyer, R.~F. 1989, Phys. Rev. D, 39, 3804}

\hi{
Schutz, B.F. 1989, Class. Quantum Grav., 6, 1761}

\hi{
Shibata, M., Nakamura, T., \& Oohara, K. 1992, Prog. Theor. Phys., 88, 1079}


\hi{
Shapiro, S.L., \& Teukolsky, S.A. 1983, {\it Black Holes, White Dwarfs, and
Neutron Stars\/} (New York: Wiley)}

\hi{
Spyromilio, J. 1994, MNRAS, 266, L61}

\hi{
Tassoul, J.-L. 1978, {\it Theory of Rotating Stars\/}
(Princeton: Princeton University Press)}

\hi{
Thompson, C., \& Duncan, R.C. 1993, ApJ, 408, 194}

\hi{
Thorne, K.S. 1987, in {\it 300 Years of Gravitation\/}, ed.
S.W. Hawking \& W. Israel (Cambridge: Cambridge Univ. Press)}

\hi{
\bysame in {\it Recent Advances in General Relativity\/},
ed. A.I. Janis \& J.R.~Porter (Birkhauser: Boston)}

\hi{
Trammell, S.R., Hines, D.C., \& Wheeler, J.C. 1993, ApJ, 414, L21}

\hi{
van den Burgh, S., \& Tammann, G.A. 1991, ARAA, 29, 363}

\hi{
van den Heuvel, E.P.J. 1987, in {\it The Origin and Evolution of Neutron
Stars\/} (IAU Symp.~125), ed. D. Helfand \& J. Huang
(Dordrecht: Reidel), p393}

\hi{
Wagoner, R. 1984, ApJ, 278, 345}

\hi{
Williams, H., \& Tohline, J.E. 1988, APJ, 334, 449}

\hi{
Yamada, S., \& Sato, K. 1993, Preprint UTAP-154/93}

\vfil\eject
\centerline{\bf Figure Captions}
\vskip 0.3truecm

\noindent
{\bf FIG.~1}.---
The secular instability growth time due to gravitational radiation
as a function of the ratio $\beta=T/|W|$ for a neutron star
modeled as a compressible Maclaurin spheroid, with $M=1.4M_\odot$
and $R_o=10$ km.
The solid line corresponds to the bar-mode ($m=2$) instability for
polytropic index $n=0$,
the dotted line for $n=0.5$, and the short-dashed
line for $n=1$. The long-dashed line is  for the $m=3$ mode and $n=0$.

\noindent
{\bf FIG.~2}.---
Secular evolution tracks of a Riemann-S ellipsoid with $n=1$ driven by
gravitational radiation reaction. The energy of an ellipsoid as a function of
the axis ratio $a_2/a_1$ is shown along various equilibrium sequences.
Dedekind-like sequences ($|\zeta|>2|\Omega|$) are shown on the left panel,
Jacobi-like ($|\zeta|<2|\Omega|$) on the right panel.
The thick solid curve corresponds to the Dedekind and Jacobi
sequences, the thick vertical line corresponds to the Maclaurin sequence,
while the other lines correspond to constant-$\cC$ sequences:
$\bar \cC=\cC/(M^3R_o)^{1/2}=-0.4$ (dotted line),
$\bar\cC=-0.32$ (dashed line), and $\bar\cC=-0.25$ (long dashed line).
The solid round dot marks the point of bifurcation.
The Dedekind-like panel is divided into region I ($\Omega>0,~\zeta>0$)
and region IV ($\Omega<0,~\zeta>0$) by the Dedekind line (the solid curve),
while the Jacobi-like panel is divided by the Jacobi line into region II
($\Omega>0,~\zeta>0$) and region III ($\Omega>0,~\zeta<0$).


\noindent
{\bf FIG.~3}.---
The total energy radiated in gravity wave during the evolution of a star
from a Maclaurin spheroid to a Dedekind ellipsoid as a function of
the initial $T/|W|$.
The solid line is for $n=0$, the dotted line for $n=0.5$, and the dashed
line for $n=1$.

\noindent
{\bf FIG.~4}.---
The gravitational waves emitted by a secularly unstable
neutron star, evolving from a Maclaurin spheroid toward a Dedekind ellipsoid.
(a) shows the gravitational wave amplitude as a function of the frequency;
(b) the number of cycles spent near $f$, and (c) the timescale
of the evolution. Here $M=1.4M_\odot$, $R_o=10$ km and $D$ is the distance to
the star. The solid lines correspond to the initial $\beta=0.24$ and
the dotted lines $\beta=0.2$, both for $n=1$;
the dashed lines correspond to $\beta=0.24$ and
the long-dashed lines $\beta=0.2$, both for $n=0.5$.

\noindent
{\bf FIG.~5}.---
(a) The maximum frequency $f_{max}$ of the gravitational wave
radiated during the evolution of a neutron star
from unstable Maclaurin spheroid
to stable Dedekind ellipsoid as a function of the initial
$T/|W|$. The solid line is for $n=0$, the dotted line for
$n=0.5$, and the dashed line for $n=1$.
(b) Twice of the rotational frequency of the initial Maclaurin
spheroid. Here $M=1.4M_\odot$ and $R_o=10$ km.

\noindent
{\bf FIG.~6}.---
Comparison between the characteristic amplitude $h_c$ of gravitational
waves emitted during the secular evolution of a nonaxisymmetric neutron star
and the rms noise $h_{rms}$ in LIGO's advanced detector.
The heavy lines in the center of the figure correspond to the evolution
shown in Fig.~4, while those in the right upper corner correspond to
the evolution shown in Fig.~7 (but here only the solid and
dotted lines are shown for clarity).

\noindent
{\bf FIG.~7}.---
Gravitational radiation from the evolution of Jacobi-like ellipsoids
(regions II and III in Fig.~2). The quantities shown are labeled
as in Fig.~4. The polytropic index is $n=1$, while
$M=1.4M_\odot$ and $R_o=10$ km.
The solid lines and the dotted lines start from a Jacobi ellipsoid,
end in a Maclaurin spheroid with $\beta=0.12$ and $\beta=0.135$.
The short-dashed lines correspond to
evolution toward a Maclaurin with $\beta=0.24$,
and the long-dashed lines to $\beta=0.02$.

\noindent
{\bf FIG.~8}.---
Terminal frequency of the Jacobi-like evolution as a function of
$T/|W|$ of the final Maclaurin spheroid.
The vertical line marks the bifurcation point.

\noindent
{\bf FIG.~9}.---
The evolution of Dedekind-like ellipsoids
(corresponding to region IV in Fig.~2)
emitting gravitational waves which carry negative angular momentum.
(a) shows the equilibrium energy,
(b) the angular momentum, (c) the axis ratio, and (d) the gravitational
wave amplitude.
$\Omega$ is the angular velocity of the ellipsoidal figure,
so that the frequency of the wave is $|\Omega|/\pi$. The solid lines
correspond to $\bar \cC=\cC/(M^3R_o)^{1/2}=-0.237$, the terminal configuration
is a Maclaurin spheroid with $\beta=0.1$; the dotted lines correspond
to $\bar \cC=-0.321$, for which the final state is a Dedekind ellipsoid
with $\Omega=0$. The polytropic index is $n=1$.

\noindent
{\bf FIG.~10}.---
Competition of Viscosity and gravitational radiation
on the stability of a Maclaurin
spheroid for $n=0$ (solid line), $n=0.5$ (dotted line),
$n=1$ (short-dashed line) and $n=1.5$ (long-dashed line).

\end